\documentclass{article}

\usepackage{arxiv}

\usepackage[utf8]{inputenc} 
\usepackage[T1]{fontenc}    
\usepackage{hyperref}       
\usepackage{url}            
\usepackage{booktabs}       
\usepackage{amsfonts}       
\usepackage{nicefrac}       
\usepackage{microtype}      
\usepackage{lipsum}

\usepackage{relsize}
\usepackage{verbatim}
\usepackage{graphicx}
\graphicspath{{Figures/}}
\usepackage{dcolumn}
\usepackage{bm}
\usepackage{subcaption}
\usepackage{float}

\usepackage{amsmath,amssymb,amsfonts}
\usepackage{graphics}
\usepackage{color}
\usepackage{xcolor}
\definecolor{rev1}{rgb}{0,0,0}

\usepackage{cite}

\usepackage[ruled,vlined]{algorithm2e}
\title{Forward Sensitivity Analysis of the FitzHugh-Nagumo System: Parameter Estimation}

\author{
 Shady E. Ahmed \\
  School of Mechanical \& Aerospace Engineering,\\
  Oklahoma State University,\\
  Stillwater, OK 74078, USA.\\
  \texttt{shady.ahmed@okstate.edu}
\And
 Omer San \\
  School of Mechanical \& Aerospace Engineering,\\
  Oklahoma State University,\\
  Stillwater, OK 74078, USA.\\
  \texttt{osan@okstate.edu} 
\And
 Sivaramakrishnan Lakshmivarahan \\
  School of Computer Science,\\
  University of Oklahoma,\\
  Norman, Oklahoma - 73019, USA.\\
  \texttt{varahan@ou.edu} \\
}

\begin{document}
\maketitle

\begin{abstract}
The FitzHugh-Nagumo (FHN) model, from computational neuroscience, has attracted attention in nonlinear dynamics studies as it describes the behavior of excitable systems and exhibits interesting bifurcation properties. The accurate estimation of the model parameters is vital to understand how the solution trajectory evolves in time. To this end, we provide a forward sensitivity method (FSM) approach to quantify the main model parameters using sparse measurement data. FSM constitutes a variational data assimilation technique which integrates model sensitivities into the process of fitting the model to the observations. We analyse the applicability of FSM to update the FHN model parameters and predict its dynamical characteristics. Furthermore, we highlight a few guidelines for observations placement to control the shape of the cost functional and improve the parameter inference iterations.
\end{abstract}


\section{Introduction} \label{sec:intro}

Dynamical systems are ubiquitous around us and in every scientific discipline. Examples from physical sciences include atmospheric and oceanic flows, heat and mass transfer, the behavior of moving objects (e.g., cars, ships, airplanes, rockets, pendulums, etc.), chemical reactions, and signal transmission. In social sciences, the population increase and distribution, human interactions, and cultural developments over centuries have been following interesting dynamical patterns. Researchers and practitioners in life sciences have also found that the application of dynamical systems theories to the bodies, organs, and cells yields significant advancement in our understanding and treatment of the body. In neurosciences, the understanding of brain performance and response to external stimulus has been critical for epilepsy prevention and treatment. Several dynamical models have been historically proposed and investigated to study and analyze the neuronal activity (e.g., see \cite{izhikevich2007dynamical}). The FitzHugh-Nagumo (FHN) equations \cite{fitzhugh1955mathematical,fitzhugh1961impulses,nagumo1962active} represent one of the very popular and simple models in the study of neuro-physiology. In addition to its utility for the modeling of biological behavior, it is considered a prototypical model in the study of nonlinear dynamics  due to its interesting characteristics such as the bifurcation properties \cite{sehgal2020numerical}.


The two-equation FHN model, describing neuronal spike discharges, can be defined as
\begin{align}
    \dfrac{\mathrm{d} v}{\mathrm{d}t} &= v - \dfrac{1}{3} v^3 - w + I, \label{eq:FNv}\\
    \tau \dfrac{\mathrm{d} w}{\mathrm{d}t} &= v+a-bw, \label{eq:FNw}
\end{align}
where $v$ defines the membrane potential, $w$ stands for a recovery variable, and $\tau$ is the time scale. $I$ represents the external input current, while $a$ and $b$ are controlling parameters. The FHN model might appear in various forms, which can be related to Eqs.~\ref{eq:FNv}--\ref{eq:FNw} by a set of changes of variables and coordinate transformations. It describes the dynamics of excitable systems which can be observed in various natural systems such as neuronal dynamics, electrocardiology, chemical reactions, and climate dynamics. However, the parameters in the FHN model are difficult to be computed directly in a real-world experimentation and the estimation of these parameters has gained the interest of a lot of researchers in physiological sciences. We shall see in the following discussions that the specification of the model parameters is crucial for the prediction of the system's behavior. For instance, the system can either converge to a stable fixed point or exhibit a limit cycle. Thus, the knowledge of such parameters can be very useful for diagnostic as well as prediction purposes, and the objective of the current study is to estimate the model's parameters from a few (possibly noisy) measurements of the system's state.

The parameter estimation framework for FHN model can be generally formulated via standard techniques such as simulated annealing, genetic algorithms, differential evolution, and Kalman filtering extensions. Besides, the known model's structure and characteristics can be utilized to customize an algorithm to estimate the parameters of the respective model. For example, the time-scale separation in the FHN model has been exploited to infer the model's parameters \cite{faghih2010fitzhugh}. Che et al. \cite{che2012parameter} solved the parameter estimation problem by deriving a second order differential equation for the membrane potential, being the observed quantity. A least-squares based regression was then applied and equipped by a wavelet denoising technique to reduce the effect of noise contamination. Geng et al. \cite{geng2020expectation} applied an expectation maximization based algorithm to identify generic FHN model parameters and estimate the variance of the interfering Gaussian noise. Jensen et al. \cite{jensen2012markov} applied a Markov chain Monte Carlo method to infer the parameters in a stochastic FHN model, constructed by adding a noise term governed by a Brownian motion. Melnykova \cite{melnykova2020parametric} proposed a contrast estimator technique to infer the model's parameters in the asymptotic setting. 

In the present study, we utilize a variational data assimilation technique, namely the forward sensitivity method (FSM) \cite{lakshmivarahan2010forward,lakshmivarahan2017forecast}, to identify the correct parameter values. The inherent sensitivity analysis reveals the relative dependence of the cost functional, defined by the discrepancy between the identified model's predictions and the actual observations, onto the respective parameters. We also investigate the effect of observation placement instants on the shape of the cost functional and the corresponding sensitivities. We finally highlight measurement collection guidelines that potentially improve the parameter inference iterations.

\section{Parameter Estimation Framework} \label{sec:fsm}
The FHN model can be described as
\begin{equation}
    \dot{\mathbf{x}} = f(\mathbf{x},\boldsymbol{\alpha}),
\end{equation}
where $\mathbf{x}=[v(t), w(t)]^T$ denotes the system's state, $\boldsymbol{\alpha} = [a,b]^T$ is the model's parameters, and $f$ represents the continuous-time dynamics of the FHN model (i.e., $f(\mathbf{x};\alpha) = [v - \dfrac{1}{3} v^3 - w + I, (v+a-bw)/\tau]^T$). Assuming the the model $f$ is continuously differentiable in its arguments (i.e., $\mathbf{x}$ and $\boldsymbol{\alpha}$), its Jacobians with respect to the state $\mathbf{x}$ and the parameter $\boldsymbol{\alpha}$ can be defined as below
\begin{align}
    Df_{\mathbf{x}} = \begin{bmatrix} 1-v^2& -1 \\  1/\tau & -b/\tau \end{bmatrix}, \qquad 
    Df_{\boldsymbol{\alpha}} = \begin{bmatrix} 0& 0 \\  1/\tau & -w/\tau \end{bmatrix},
\end{align}
where $Df_{\mathbf{x}}$ and $Df_{\boldsymbol{\alpha}}$ define the model's sensitivity with respect to the state $\mathbf{x}$ and the parameters $\boldsymbol{\alpha}$, respectively. 

\subsection{Forward sensitivities}
Using a suitable temporal integration scheme, the FHN can be rewritten in a discrete-time form as follows,
\begin{equation}
    \mathbf{x}({k+1}) = \mathbf{M}(\mathbf{x}(k), \boldsymbol{\alpha}), \label{eq:disc}
\end{equation}
where $\mathbf{x}(k)=[v(t_k), w(t_k)]^T \in \mathbb{R}^2$ defines the system's state at time $t_k$, and $\mathbf{M}: \mathbb{R}^2 \times \mathbb{R}^2 \to \mathbb{R}^2$ represents the one-step state transition map. Thus, the following discrete-time Jacobians can be computed,
\begin{align}
    \mathbf{D} \mathbf{M}_{\mathbf{x}} (k) = \bigg[ \dfrac{\partial M_i(\mathbf{x},\boldsymbol{\alpha}) }{\partial x_j} \bigg]_{\mathbf{x} = \mathbf{x}(k)}, \qquad  
    \mathbf{D} \mathbf{M}_{\boldsymbol{\alpha}} (k) = \bigg[ \dfrac{\partial M_i(\mathbf{x},\boldsymbol{\alpha}) }{\partial \alpha_j} \bigg]_{\mathbf{x} = \mathbf{x}(k)}.
\end{align}
Furthermore, we define the sensitivity of the model forecast at any time $t_k$ with respect to the model's parameters as follows,
\begin{equation}
    \mathbf{V}(k) = \bigg[ \dfrac{\partial x_i(k)}{\partial \alpha_j} \bigg] \in \mathbb{R}^{2\times 2}.
\end{equation}
Equation~\ref{eq:disc} can be used to evaluate the forward sensitivity matrices at different times in a recursive way as
\begin{equation}
    \mathbf{V}(k+1) =  \mathbf{D} \mathbf{M}_{\mathbf{x}} (k) \mathbf{V}(k) +  \mathbf{D} \mathbf{M}_{\boldsymbol{\alpha}} (k),
\end{equation}
with $\mathbf{V}(0)=\mathbf{0}$ since the initial condition $\mathbf{x}(0)$ is independent of the model's parameters $\boldsymbol{\alpha}$.

\subsection{Forecast error}
Assuming $\mathbf{z}(k) \in \mathbb{R}^m$ to be the vector of measurements at time $t_k$, given by
\begin{equation}
    \mathbf{z}(k) = \mathbf{h}(\bar{\mathbf{x}}(k)) + \boldsymbol{\xi}(k),
\end{equation}
where $\mathbf{h}: \mathbb{R}^2 \to \mathbb{R}^m$ defining the observational operator that relates the model space to the observation space, and $\bar{\mathbf{x}}$ defines the \emph{true} system's state while $\boldsymbol{\xi}$ denotes the measurement noise. For simplicity, we suppose that we directly measure the system's state (i.e., $\mathbf{h}(\mathbf{x}(k)) = \mathbf{x}(k)$). We also assume that $\boldsymbol{\xi}$ is a white Gaussian noise with zero mean and a covariance matrix $\mathbf{R}$ (i.e., $ \boldsymbol{\xi}(k) = {\mathcal{N}}(\mathbf{0},\mathbf{R}(k))$).

We define the difference between the model forecast and measurements as $\mathbf{e}(k) = \mathbf{z}(k) - \mathbf{h}(\mathbf{x}(k))$, which is called the innovation or forecast error (computed in the observation space). With the assumption that the dynamical model is perfect (i.e., correctly encapsulates all the relevant processes) and the initial condition $\mathbf{x}(0)$ is known, then the deterministic part of the forecast error can be attributed to the inaccuracy of the model's parameters values, defined as $\delta \boldsymbol{\alpha} = \bar{\boldsymbol{\alpha}} - \boldsymbol{\alpha}$, where $\bar{\boldsymbol{\alpha}}$ denotes the true values of the parameters. Thus, we can define a cost functional $J:\mathbb{R}^2 \to \mathbb{R}$ as 
\begin{equation}
    J(\boldsymbol{\alpha}) = \sum_{k=1}^{N}  \dfrac{1}{2} \| \mathbf{e}(k) \|^2_{\mathbf{R}^{-1}(k)} = \sum_{k=1}^{N} \dfrac{1}{2} \mathbf{e}(k)^T \mathbf{R}^{-1}(k) \mathbf{e}(k),  \label{eq:cost} 
\end{equation}
where $N$ is the number of measurement instants. The minimization of the cost function $J$ can be solved as a strong constrained problem with the standard Lagrangian multiplier method, resulting in the adjoint framework. Alternatively, we utilize the forward sensitivity matrices to evaluate an optimal estimate for the parameters $\boldsymbol{\alpha}$. Let $\delta \mathbf{x}(k) = \bar{\mathbf{x}}(k) - \mathbf{x}(k)$ be the difference between the model's forecast and the true state, with $\delta \mathbf{x}(0)=0$ since the initial conditions are perfectly known. With first order Taylor expansions of $\mathbf{e}(k)$ and $\delta \mathbf{x}(k)$, the following expressions can be defined,
\begin{align}
    \mathbf{e}(k) = \mathbf{D}\mathbf{h}(k) \delta \mathbf{x}(k) ,\qquad
    \delta \mathbf{x}(k) = \mathbf{V}(k) \delta \boldsymbol{\alpha},
\end{align}
where $\mathbf{D}\mathbf{h}$ is the Jacobian of the observational operator $\mathbf{h}$. Therefore, the forecast error can be related to the correction to the model's parameters as $\mathbf{e}(k) = \mathbf{D}\mathbf{h}(k) \mathbf{V}(k) \delta \boldsymbol{\alpha}$. Since we assume that $\mathbf{h}(\mathbf{x}(k)) = \mathbf{x}(k)$, we deduce that $\mathbf{D}\mathbf{h}$ reduces to the identity matrix. The previous forecast error formulation can be written for all $N$ time instants at which observations become available, and the following linear equation is obtained,
\begin{equation} \label{eq:E_dc}
    \mathbf{H} \delta \boldsymbol{\alpha} = \mathbf{e}_F,
\end{equation}
where the matrix $\mathbf{H} \in \mathbb{R}^{Nm \times 2}$ and the vector $\mathbf{e}_F \in \mathbb{R}^{N m}$ are defined as follows,
\begin{equation} \label{eq:H_e_def}
    \mathbf{H} = \begin{bmatrix}     
    \mathbf{D}\mathbf{h}(1) \mathbf{V}(1)  \\
    \mathbf{D}\mathbf{h}(2) \mathbf{V}(2)  \\
    \vdots \\
    \mathbf{D}\mathbf{h}(N) \mathbf{V}(N)    
    \end{bmatrix}, \qquad
    \mathbf{e}_F = \begin{bmatrix}
    \mathbf{e}(1) \\
    \mathbf{e}(2) \\
    \vdots \\
    \mathbf{e}(N)
    \end{bmatrix}.
\end{equation}
The inverse problem can be solved in a weighted least squares sense to find an optimal correction vector $\delta \boldsymbol{\alpha}$, with $\mathbf{R}^{-1}$ as a weighting matrix, where $\mathbf{R}$ is a an $Nm \times Nm$ block-diagonal matrix with $\mathbf{R}(k)$ being its $k$-th diagonal block. We assume that $\mathbf{R}$ is a diagonal matrix defined as $\mathbf{R} = \sigma^2 \mathbf{I}_{Nm}$, where $\mathbf{I}_{Nm}$ is the $Nm \times Nm$ identity matrix. Then, the solution to Eq.~\ref{eq:E_dc} can be written as
\begin{equation}\label{eq:LSsolve}
\delta \boldsymbol{\alpha} = \left( \mathbf{H}^T \mathbf{R}^{-1} \mathbf{H} \right)^{-1} \mathbf{H}^T \mathbf{R}^{-1} \mathbf{e}_{F}.
\end{equation}

\subsection{Placement of observations using forward sensitivity} \label{sec:placement}
In order to select the time instants at which measurement data are collected, we relate the cost functional given in Eq.~\ref{eq:cost} to the forward sensitivity matrix $\mathbf{V}(k)$. This is based on the method proposed by Lakshmivarahan et al. \cite{lakshmivarahan2020controlling} to control the shape of the cost functional and keep its gradient away from zero to accelerate the convergence. By substituting $\mathbf{e}(k) = \mathbf{D}\mathbf{h}(k) \mathbf{V}(k) \delta \boldsymbol{\alpha}$ into Eq.~\ref{eq:cost}, we get the following,
\begin{equation}
   J(\boldsymbol{\alpha}) = \sum_{k=1}^{N} \dfrac{1}{2} \delta \boldsymbol{\alpha}^T \bigg(\mathbf{E}(k)^T \mathbf{R}^{-1}(k) \mathbf{E}(k) \bigg) \delta \boldsymbol{\alpha} = \sum_{k=1}^{N} \dfrac{1}{2} \delta \boldsymbol{\alpha}^T \mathbf{G}(k) \delta \boldsymbol{\alpha},
\end{equation}
where $\mathbf{E}(k)=\mathbf{D}\mathbf{h}(k) \mathbf{V}(k)$ and $\mathbf{G}(k) = \mathbf{E}(k)^T \mathbf{R}^{-1}(k) \mathbf{E}(k)$. We note that $\mathbf{G}(k)$ is called the observability Gramian. The gradient of the cost functional with respect to the parameter vector $\boldsymbol{\alpha}$ can be written as below
\begin{align}
    \nabla_{\boldsymbol{\alpha}} J(\boldsymbol{\alpha}) &= \sum_{k=1}^{N} - \mathbf{G}(k) \delta \boldsymbol{\alpha}, \label{eq:nabla}
\end{align}
which relates the gradient of the cost functional and the parameterization error/correction. From Eq.~\ref{eq:nabla}, a necessary condition for the minimization of the cost functional is that $\mathbf{G}(k)$ is positive definite. For the case considered here, $\mathbf{D}\mathbf{h}(k) = \mathbf{I}$ and $\mathbf{R}^{-1}(k) =\dfrac{1}{\sigma^2} \mathbf{I}$. Thus, $\mathbf{G}(k) = \dfrac{1}{\sigma^2} \mathbf{V}(k)^T \mathbf{V}(k)$, where $\mathbf{V}(k) = \begin{bmatrix} V_{11} & V_{12} \\ V_{21} & V_{22} \end{bmatrix}$. Therefore, one way to guarantee that the gradient of the cost functional does not hit zero and improve the convergence is to select the measurement instants in such a way that the diagonal entries (i.e., $ V^2_{11} + V^2_{21}$ and $V^2_{12} + V^2_{22}$) are as large as possible.



\section{Results and Discussions} \label{sec:res}
We analyze the capability of the forward sensitivity approach to identify the FHN model's parameters. In particular, we study an arbitrary case where the true parameters values are $a=0.15$ and $b=0.35$. Initial conditions of $(v(0),w(0)) = (0.0,1.0)$ are considered and the fourth order Runge-Kutta scheme is applied for time integration with a time step of $\Delta t = 0.1$, time scale $\tau=10$, and a maximum time of $t_m= 100$.  We assume that the measurements are collected every $200$ time steps, corrupted by an additive Gaussian noise with a zero mean and a standard deviation of $\sigma = 0.1$. 
\subsection{Fixed input}
\begin{figure}[ht!]	
	\centering
	\includegraphics[width=0.90\linewidth]{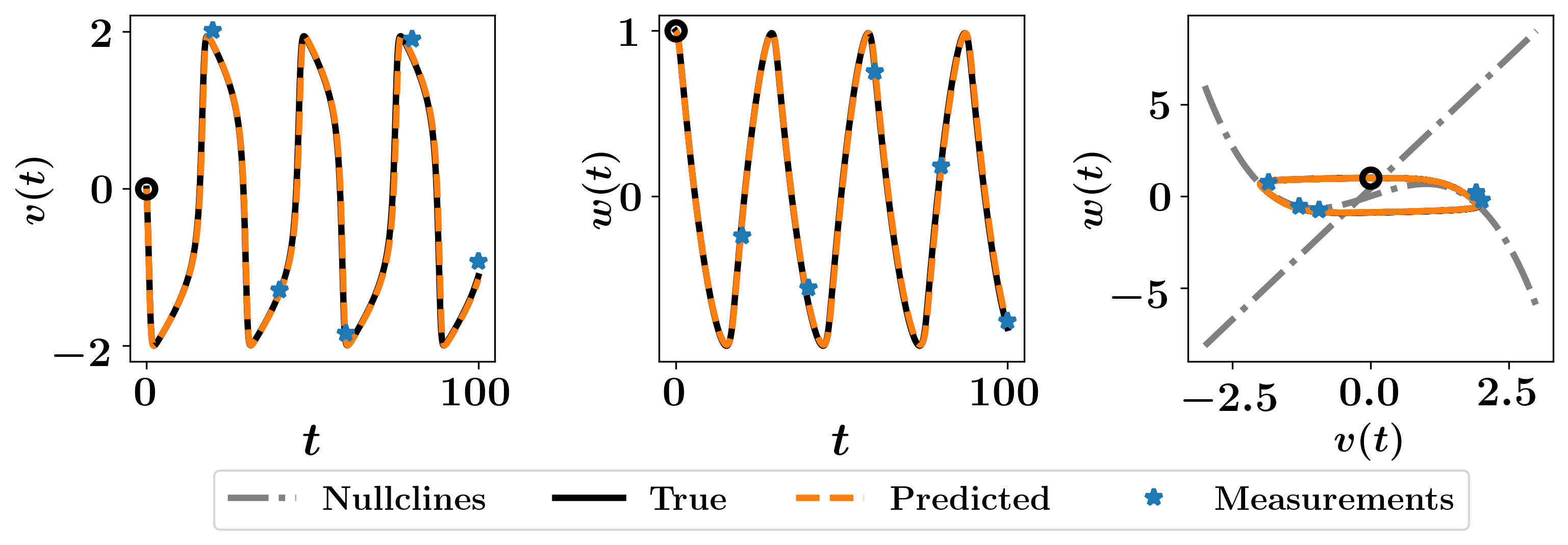}
	\caption{Results for $a=0.15$, $b=0.35$, and $I=0$, with measurements every 200 time steps and $\sigma=0.1$. Estimated parameters are $a=0.159$ and $b=0.364$.}\label{fig:I0}		
\end{figure}

As a first investigation, we study the case with zero input (i.e., $I=0$). This corresponds to a fixed point of $(v^*,w^*)=(-0.229,-0.225)$ with a model Jacobian of $\begin{bmatrix} 0.948 & -1\\ 0.1 & -0.035\end{bmatrix}$. The eigenvalues of this matrix are $\lambda_{1} = 0.832, \lambda_2= 0.080$, implying unsteady equilibrium points. However, a Lyapunov function analysis reveals that the solution of this system is bounded and exhibits an attractive limit cycle \cite{ringkvist2006dynamical,kaumann1983uniqueness,hadeler1976generation,treskov1996existence,ringkvist2009dynamical}. In Figure~\ref{fig:I0}, we plot the time evolution of the membrane potential, $v$, and the recovery variable, $w$, for the true system compared to the case with the inferred parameters values. Starting from a prior guess of $a=0.2$ and $b=0.2$ to initiate the FSM iterations, a parameterization of $a=0.159$ and $b=0.364$ is identified, very close to the true values. Thus, we can see that the adopted FSM approach is adequately capable of assimilating these noisy data to estimate the model's parameters for this case.

\begin{figure}[ht!]	
	\centering
	\includegraphics[width=0.90\linewidth]{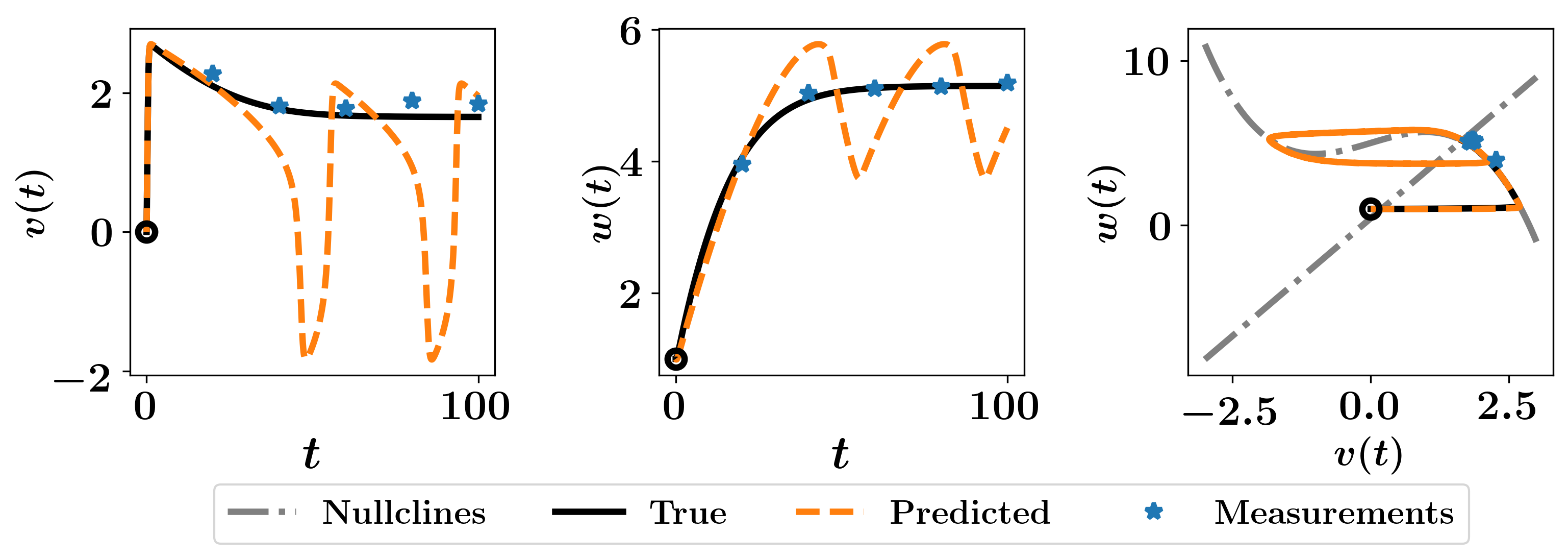}
	\caption{Results for $a=0.15$, $b=0.35$, and $I=5$, with measurements every 200 time steps and $\sigma=0.1$. Estimated parameters are $a=-0.930$ and $b=-0.021$, starting from an initial guess of $(0.2,0.2)$.}
	\label{fig:I52}
\end{figure}

A second testing situation is to apply a constant input of $I=5$, with the same parameters values as before. We find that this case corresponds to a stable fixed point. In other words, the solution trajectory converges to the equilibrium point (which is $(v^*,w^*) = (1.652,5.149)$) and resides there. We apply the same procedure to estimate the model's parameters starting with an initial guess of $a=0.2$ and $b=0.2$. The plots in Figure~\ref{fig:I52} show that the iterative algorithm fails to correctly approximate the parameters values and produces a periodic solution, instead. To understand this, we compute the fixed point and the eigenvalues of the corresponding model's Jacobian. We find that with $I=0$ (the previous case), both the true values $(a,b)=(0.15,0.35)$ and the initial guess $(a,b)=(0.2,0.2)$ induce a periodic limit cycle. On the other hand, for $I=5$, the true parameter values correspond to a stable fixed point, while the initial guess still yields a cyclic behavior. Therefore, the estimation process should cross the bifurcation points in order to predict the correct parameterization, which is a common problem in parameter estimation frameworks.

\begin{figure}[ht!]	
	\centering
	\includegraphics[width=0.90\linewidth]{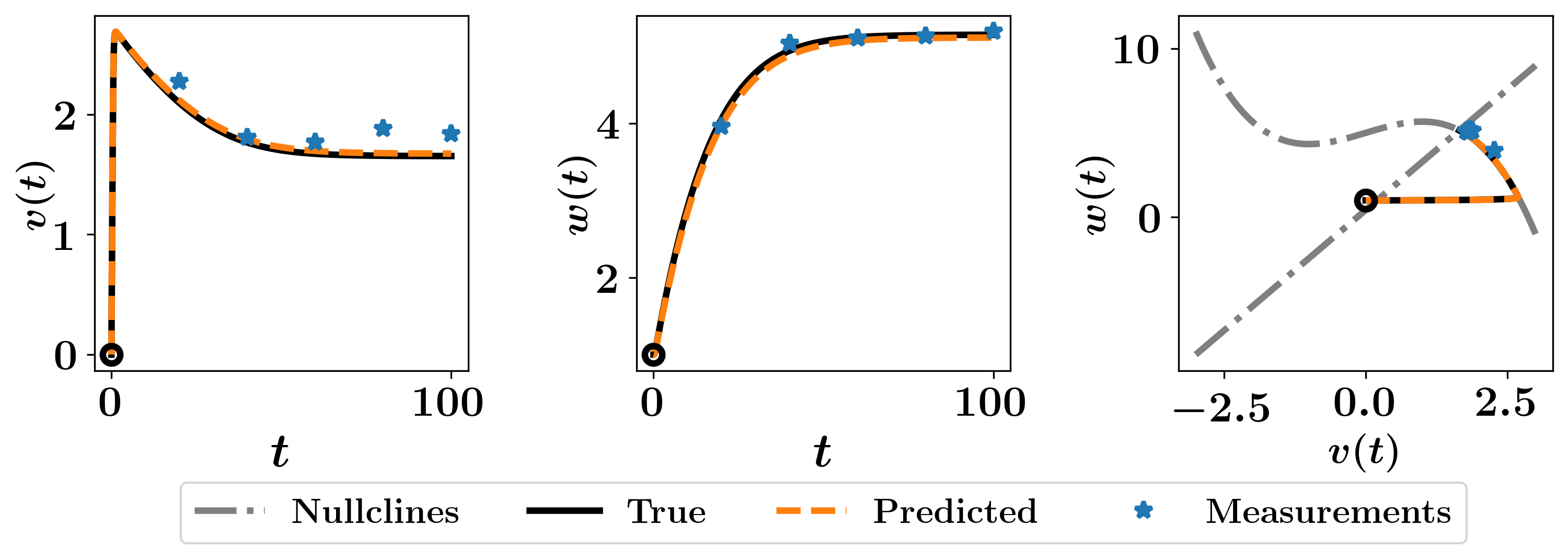}
	\caption{Results for $a=0.15$, $b=0.35$, and $I=5$, with measurements every 200 time steps and $\sigma=0.1$. Estimated parameters are $a=0.0344$ and $b=0.334$, starting from an initial guess of $(0.5,0.5)$.}
	\label{fig:I55}	
\end{figure}

In order to mitigate this issue, prior information about the regime of the solution trajectory can be utilized to make an intelligent guess. For instance, an initial guess of  $(a,b)=(0.5,0.5)$ with $I=5$ yields a stable fixed point, and hence can be chosen as an alternative starting point. Results are presented in Figure~\ref{fig:I55}, where we can see that both the true and predicted trajectories converge to the equilibrium state. However, the estimated parameters values ($a=0.0344$ and $b=0.334$) are slightly far from the true ones.

\begin{figure}[ht!]	
	\centering
	\includegraphics[width=0.90\linewidth]{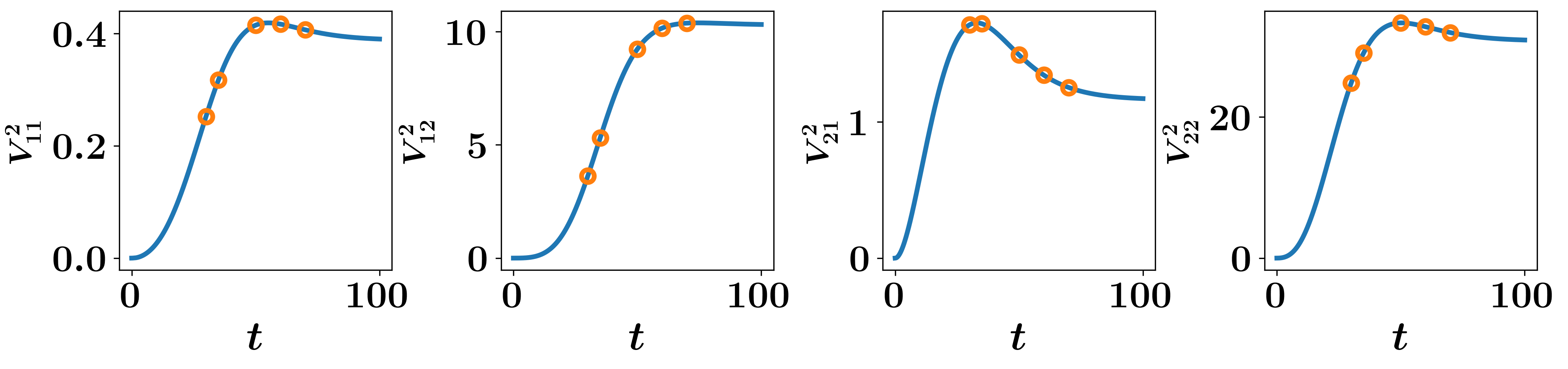}
	\caption{Forward sensitivities for $a=0.15$, $b=0.35$, and $I=5$. Selected observations instants are denoted with orange circles.}\label{fig:sens}	
\end{figure}

\begin{figure}[ht!]	
	\centering
	\includegraphics[width=0.90\linewidth]{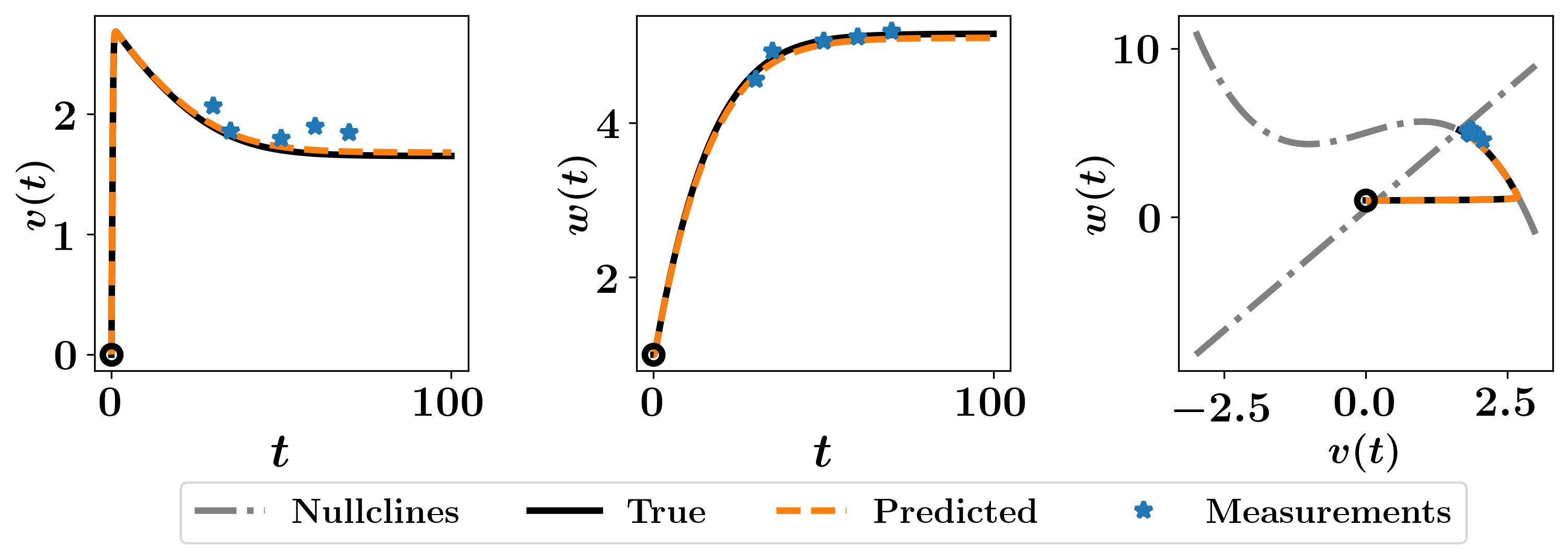}
	\caption{Results for $a=0.15$, $b=0.35$, and $I=5$, with measurements instants selected based on the forward sensitivity criteria. Estimated parameters are $a=0.128$ and $b=0.355$, starting from an initial guess of $(0.5,0.5)$.}\label{fig:I5s}
\end{figure}

In order to explore the effect of the measurements on the forward sensitivities, we plot the variation of $V^2_{ij}$ for $i,j\in\{1,2\}$ with time in Figure~\ref{fig:sens}. We observe that the initial period has the least influence on the forward sensitivities, while the measurements around and after $t=50$ have the largest effects. Therefore, we redistribute our measurement instants based on the approach described in Section~\ref{sec:placement}. In particular, we collect data at $t \in \{30, 35, 50, 60, 70\}$ and apply the FSM framework to estimate the model's parameter. Starting from an initial guess of $(a,b)=(0.5,0.5)$, a parameterization of $(a,b)=(0.128,0.355)$ is estimated, showing significant improvement with respect to the case with equispaced measurement signals. Results are shown in Figure~\ref{fig:I5s} for the true and predicted trajectories. We can also notice that the optimized measurements are concentrated towards the equilibrium state. 

\subsection{Varying input}
\begin{figure}[ht!]	
	\centering
	\includegraphics[width=0.90\linewidth]{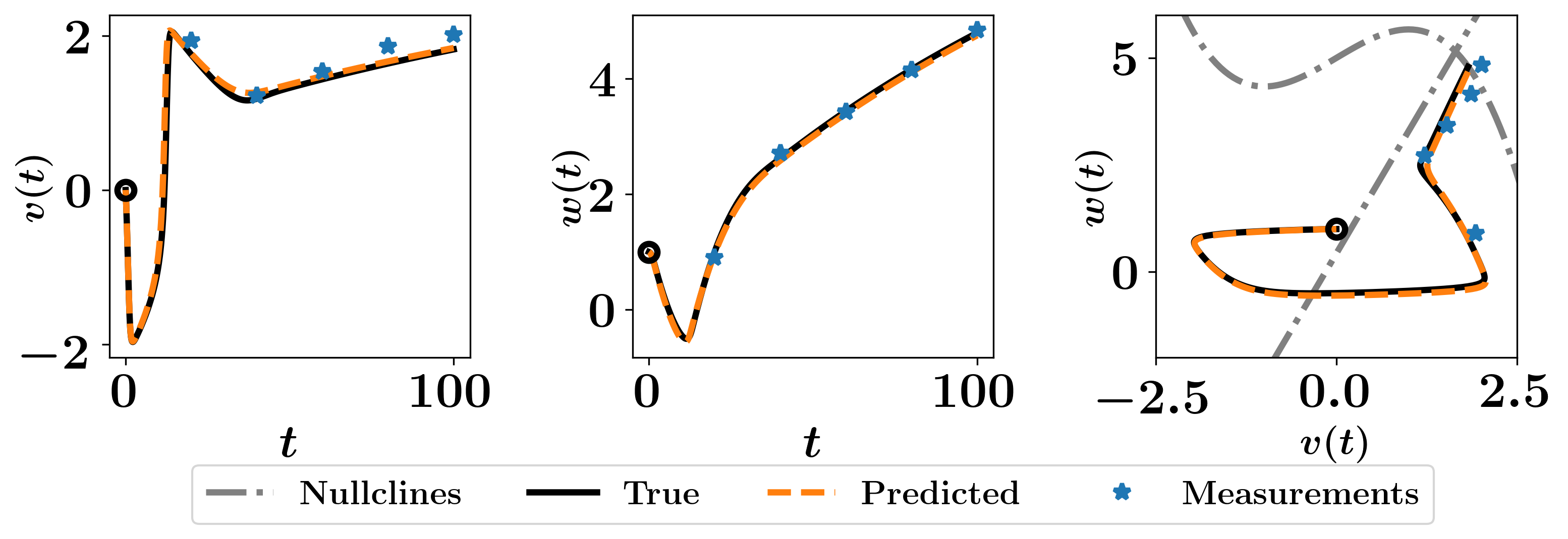}
	\caption{Results for $a=0.15$, $b=0.35$, and varying $I$, with measurements every 200 time steps and $\sigma=0.1$. Estimated parameters are $a=0.023$ and $b=0.330$, starting from an initial guess of $(0.2,0.2)$.}
	\label{fig:Ivary}	
\end{figure}

Finally, we vary the input excitation as $I=5t/t_m$ (i.e., linearly increasing from $0$ to $5$). This corresponds to a moving fixed point, beginning with a cyclic trajectory and followed by a convergence to the stable equilibria. Parameter estimation results for equidistant measurement instants are depicted in Figure~\ref{fig:Ivary} beginning from an initial guess of $(a,b)=(0.2,0.2)$. We find that the predicted trajectory sufficiently match the true one, but the estimated parameters are not very accurate.

\begin{figure}[ht!]	
	\centering
	\includegraphics[width=0.90\linewidth]{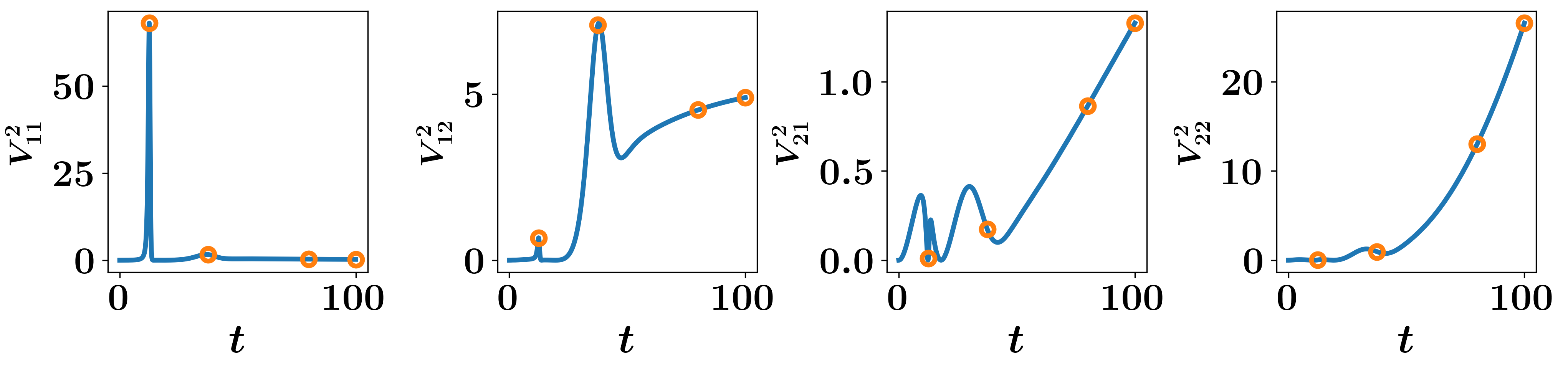}
	\caption{Sensitivities for $a=0.15$, $b=0.35$, and varying $I$. Selected observations instants are denoted with orange circles.}
	\label{fig:vsens}	
\end{figure}

\begin{figure}[ht!]	
	\centering
	\includegraphics[width=0.90\linewidth]{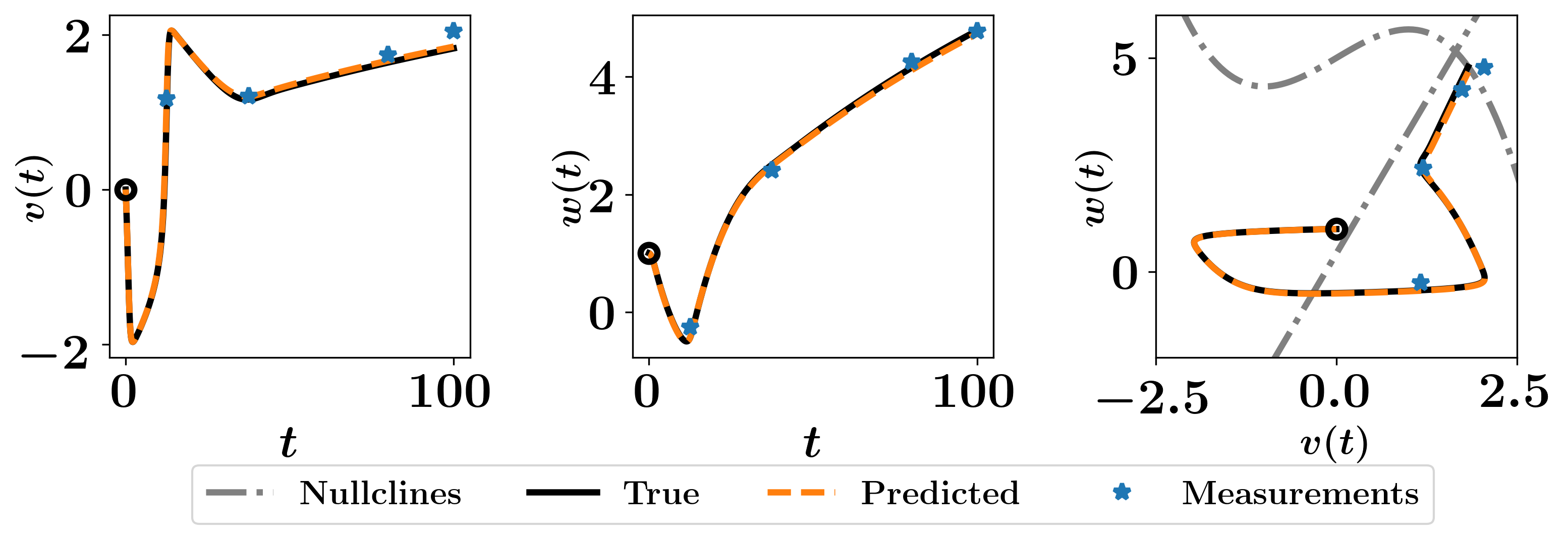}
	\caption{Results for $a=0.15$, $b=0.35$, and varying $I$, with measurements instants selected based on the forward sensitivity criteria. Estimated parameters are $a=0.128$ and $b=0.356$ starting from an initial guess of $(0.2,0.2)$.}
	\label{fig:Ivarys}	
\end{figure}

We then investigate the effects of observation times on the forward sensitivities of the model predictions. We find a spike in the sensitivity of $v$ predictions with respect to the $v$ measurements around $t=12.5$. We also see a relatively large dependence on the $w$ measurements about $t=37.5$. On the other hand, the $w$ predictions show an increasing sensitivity on either $v$ or $w$ measurements at final times. Therefore, we reallocate our observation times to capture these trends as demonstrated in Figure~\ref{fig:vsens}. Results based on this enhanced parameter estimation methodology are described in Figure~\ref{fig:Ivarys}, where the approximated parameters values ($a=0.128$, and $b=0.356$) are closer to the true values. 

\section{Concluding Remarks} \label{sec:conc}
We put forth a forward sensitivity analysis for the FitzHugh-Nagumo (FHN) system to infer the model's parameterization from sparse observations. The approach relies on the investigation of the forward sensitivity matrices that encapsulates the temporal dependence of model's predictions onto its parameters. The presented methodology shows substantial success in assimilating noisy observational data to identify the unknown parameters. We find that the convergence of the predicted parameters to the true values relatively depends on the first guess used to initialize the algorithm. In particular, the initial guess has to yield equilibrium points with similar stability characteristics to the true one. We study three test cases, including zero input, constant non-zero current, and time-dependent excitation. We also formulate measurement collection guidelines based on the relation between the cost functional and the forward sensitivity components. We demonstrate that this approach provides more accurate estimates of unknown parameters than those resulting with arbitrary measurement placements.\\

\section*{Acknowledgments}
This material is based upon work supported by the U.S. Department of Energy, Office of Science, Office of Advanced Scientific Computing Research under Award Number DE-SC0019290. O.S. gratefully acknowledges the U.S. DOE Early Career Research Program support.

Disclaimer: This report was prepared as an account of work sponsored by an agency of the United States Government. Neither the United States Government nor any agency thereof, nor any of their employees, makes any warranty, express or implied, or assumes any legal liability or responsibility for the accuracy, completeness, or usefulness of any information, apparatus, product, or process disclosed, or represents that its use would not infringe privately owned rights. Reference herein to any specific commercial product, process, or service by trade name, trademark, manufacturer, or otherwise does not necessarily constitute or imply its endorsement, recommendation, or favoring by the United States Government or any agency thereof. The views and opinions of authors expressed herein do not necessarily state or reflect those of the United States Government or any agency thereof.

\bibliographystyle{unsrt} 
\bibliography{references}

\end{document}